\documentclass[reprint, 
 amsmath,amssymb,
 aps, physrev,
]{revtex4-2}
\usepackage{mathptmx}
\usepackage{graphicx}
\usepackage{dcolumn}
\usepackage{bm}
\usepackage[colorlinks=true, linkcolor=blue, urlcolor=blue, citecolor=blue, filecolor=magenta]{hyperref}
\usepackage{multirow}
\usepackage{booktabs}
\usepackage{longtable}

\usepackage{siunitx}
\usepackage{color, soul}
\sethlcolor{yellow}
\begin{document}

\title{\textbf{Evolution of Jeans instability in nonthermal magnetized complex plasmas} 
}%

\author{Dipankar Ray\href{https://orcid.org/0009-0002-2308-7850}{\includegraphics[scale=0.08]{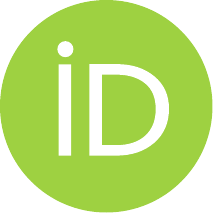}}}
\email{Contact author: dray056@tezu.ernet.in}
 \affiliation{Department of Physics, Tezpur University, Napaam, Tezpur, Sonitpur, Assam - 784028, India.}
\author{Bharati Kakad\href{https://orcid.org/0000-0001-8440-3373}{\includegraphics[scale=0.08]{orcid.pdf}}}
\affiliation{
Indian Institute of Geomagnetism, New Panvel, Navi Mumbai - 410218, India.
}
\author{Amar Kakad\href{https://orcid.org/0000-0002-1308-4245}{\includegraphics[scale=0.08]{orcid.pdf}}}
\affiliation{
Indian Institute of Geomagnetism, New Panvel, Navi Mumbai - 410218, India.
}
\author{Pralay Kumar Karmakar\href{https://orcid.org/0000-0002-3078-9247}{\includegraphics[scale=0.08]{orcid.pdf}}}%
\affiliation{Department of Physics, Tezpur University, Napaam, Tezpur, Sonitpur, Assam - 784028, India.}%
\date{\today}

\begin{abstract}
The classical Jeans (gravitational) instability framework provides a foundational basis for understanding self-gravitational collapse and structure formation in interstellar environments. Magnetic fields and nonthermal (non-Maxwellian) particle populations are both common in interstellar clouds, yet their combined influence on gravitational collapse has not been explored extensively beyond linear theory. We investigate the nonlinear evolution of the Jeans instability in a magnetized nonthermal complex (dusty) plasma using linear dispersion analysis and one-dimensional fluid simulations. The inertial, negatively charged dust grains are modeled as a cold fluid coupled to a uniform, frozen-in transverse magnetic field, while the lighter electrons and ions follow $\kappa$-velocity distributions. The numerical architecture utilizes a fourth-order finite difference method for spatial gradients and a second-order leap-frog time-integration scheme. Linear analysis shows that the transverse magnetic field introduces a $\kappa$-independent compressional term that elevates the critical self-gravity threshold. In nonlinear simulations, inward gravitational contraction compresses the frozen-in magnetic field, generating outward magnetic pressure that substantially reduces the instability growth rate across all $\kappa$. While enhanced nonthermality (smaller $\kappa$) gives a higher growth rate at weak magnetic fields, this $\kappa$-dependence weakens as the field strength increases. In the strong-field regime, the growth rates of thermal and nonthermal systems nearly coincide, indicating that compressional magnetic pressure governs instability growth rate in strongly magnetized environments, whereas nonthermality remains consequential primarily in weak- to moderate-field regimes.
\end{abstract}

\maketitle
\section{Introduction}\label{sec:introduction}
Stars are the primary building blocks of galaxies. Unraveling their complex life cycle, from formation to the final evolutionary stages, is essential to gain comprehensive insights into the physical evolution and chemical enrichment of the cosmos itself~\cite{larsonPhysicsStarFormation2003,liuDominanceTurbulenceMagnetism2026}. Consequently, understanding the mechanisms of stellar genesis is of foundational significance to both astrophysics and plasma physics. Star formation is widely modeled as the gravitational fragmentation of an initially extended medium via the Jeans instability or gravitational instability~\cite{jeansStabilitySphericalNebula1901, choudhuriAstrophysicsPhysicists2010, larsonPhysicsStarFormation2003}. In its classical formulation for a neutral, isothermal gas, collapse sets in when self-gravity overcomes the stabilizing thermal pressure~\cite{jeansStabilitySphericalNebula1901}. However, the presence of charged dust grains is an inevitable feature across a wide variety of astrophysical systems, such as planetary rings, comet tails, and dense interstellar molecular clouds \cite{verheestWavesInstabilitiesDusty1996, paganiUbiquityMicrometerSizedDust2010, draineOpticalPropertiesInterstellar1984}. These micron- to sub-micron-sized solid particulates, typically composed of graphites and silicates, can acquire substantial electric charges ranging from tens to thousands of elementary charges. Thus, the star-forming environments can be more accurately described as complex (dusty) plasmas, in which massive charged grains are embedded in an electron--ion background~\cite{tsintsadzeJeansInstabilityMagnetoradiative2008, PhysRevE.49.5599, AVINASH1994470}. In such multicomponent systems, the gravitational collapse can differ substantially from the classical Jeans instability of a neutral gas.

Astrophysical dust grains are typically negatively charged and exert mutual electrostatic repulsion under compression, which acts as an additional restoring force (alongside thermal pressure) that opposes the inward gravitational contraction~\cite{PhysRevE.49.5599, AVINASH1994470, shuklaJeansInstabilitySelfGravitating2006}. Moreover, energy dissipation mechanisms such as stellar winds, supernova-driven shocks, and magnetic reconnection can accelerate the electrons and ions, producing nonthermal (non-Maxwellian) populations with prominent power-law tails at velocities exceeding the characteristic thermal speed. Satellite measurements and \textit{in situ} observations indicate that astrophysical plasmas are frequently in non-equilibrium stationary states where particle velocity distributions deviate from the standard Maxwell--Boltzmann thermal equilibrium~\cite{vasyliunasLowEnergyElectronsMagnetosphere1968,maksimovicUlyssesElectronDistributions1997,pierrardElectronVelocityDistribution1999,nichollsResolvingElectronTemperature2012}. Such nonthermal particles can screen grain potentials more effectively than thermal particles, reducing the effective Debye length and weakening electrostatic support against gravity \cite{ray_PRE_2026}. These high-energetic populations are commonly described by the $\kappa$-distribution, which has a power-law high-energy tail and reduces to the Maxwellian as the nonthermal spectral index $\kappa\to\infty$~\cite{vasyliunasLowEnergyElectronsMagnetosphere1968, pierrardKappaDistributionsTheory2010}.

Beyond the electrostatic and gravitational coupling, ambient magnetic fields are ubiquitous in cosmic environments such as star-forming clouds and protoplanetary disks. Zeeman observations of the diffuse interstellar medium and molecular-cloud envelopes reveal field strengths of a few to $\sim\!10\,\mu$G, with a median of $\approx6\,\mu$G in the cold neutral medium~\cite{crutcherMAGNETICFIELDSINTERSTELLAR2010, heilesMillenniumArecibo212005}. For perturbations propagating perpendicular to the magnetic field, magnetic pressure introduces additional stabilizing support that can suppress or halt the Jeans instability~\cite{nakanotaniNonlinearKineticSimulations2025a}. High-resolution dust polarization observations of massive star-forming regions with the Atacama Large Millimeter/submillimeter Array (ALMA) directly confirm this directional interplay, demonstrating that magnetic fields oppose gravitational infall most effectively in regions where the field orientation is nearly orthogonal to local gravity~\cite{kochPolarizationPropertiesMagnetic2018}. Several theoretical studies have similarly emphasized this stabilizing role against gravity across intermediate- to small-scale star-forming regions, including giant molecular clouds and individual prestellar cores~\cite{tsintsadzeJeansInstabilityMagnetoradiative2008,chandrasekhar1953problems,kolesnichenkoJeansInstabilityProtoplanetary2025}.

Linear normal-mode analysis provides an essential baseline for how nonthermal particles and magnetic fields modify the Jeans length and instability threshold. It is, however, limited to small-amplitude perturbations and cannot describe the subsequent gravitational runaway, localized structural sharpening, or saturation, which require numerical simulation. While three-dimensional ideal magnetohydrodynamic (MHD) simulations have been instrumental in mapping magnetic field topologies in the turbulent interstellar medium, such models frequently omit self-gravity to isolate purely kinematic effects~\cite{falceta-goncalvesStudiesRegularRandom2008}. To bridge the gap between linear theory and nonlinear dynamics, we previously developed a fluid simulation framework for the Jeans instability with $\kappa$-distributed lighter species and a cold dust fluid, showing that nonthermality shortens the characteristic collapse time while self-gravity primarily governs the growth rate~\cite{ray_PRE_2026}. However, that study was restricted to an unmagnetized regime, whereas recent magnetized kinetic simulations of the instability have typically assumed thermal species~\cite{nakanotaniNonlinearKineticSimulations2025a}. The combined influence of an ambient magnetic field and nonthermal particle statistics on nonlinear gravitational collapse therefore remains insufficiently explored. Here, we extend our framework by coupling a cold dust fluid to a frozen-in transverse magnetic field and follow the instability from its linear onset into the nonlinear regime to determine how the field strength and degree of nonthermality jointly determine the growth rate.

The remainder of this paper is organized as follows: in Sec.~\ref{sec:physical model}, we formulate the physical model along with the normalized governing equations and derive the linear dispersion relation to establish the instability criteria. In Sec.~\ref{sec:simulation setup}, the numerical scheme and simulation setup are described. The nonlinear evolution and parametric trends are presented and discussed in Sec.~\ref{sec:results and discussion}. Finally, Sec.~\ref{sec:conclusions} summarizes the findings and conclusions of this investigation.

\section{Physical Model and Linear Dispersion Analysis}\label{sec:physical model}
We consider a self-gravitating, collisionless, and magnetized tricomponent dusty plasma consisting of massive, negatively charged dust grains, along with high-energy nonthermal electrons and ions. The inertial dust grains are treated as a compressible fluid and the lighter electrons and ions are considered to follow the $\kappa$ velocity distribution. The wave propagation and gravitational collapse are restricted to a one-dimensional (1D) Cartesian axis ($x$-axis), while a uniform ambient magnetic field is applied perpendicular to the direction of propagation ($z$-axis). Physically, when matter infalls parallel to magnetic field lines, the magnetic field offers negligible resistance; conversely, high-resolution dust polarization observations of star-forming cores confirm that magnetic forces oppose gravitational collapse most effectively when the magnetic field orientation is orthogonal to the local gravitational pull~\cite{kochPolarizationPropertiesMagnetic2018}. Hence, adopting a transverse magnetic field isolates the regime of maximum magnetogravitational coupling.

\subsection{Dust fluid and field equations}
Under the action of collective electrostatic, self-gravitational, and compressional magnetic forces, the hydrodynamic transport of the dust species is governed by the 1D fluid continuity and momentum equations, respectively:
\begin{equation}
    \frac{\partial n_d}{\partial t} + \frac{\partial}{\partial x}(n_d v_d) = 0,
    \label{eq:1}
\end{equation}
\begin{equation}
    \frac{\partial v_d}{\partial t} + v_d\frac{\partial v_d}{\partial x} = \frac{Z_d e}{m_d} \frac{\partial \phi}{\partial x} - \frac{\partial \psi}{\partial x} - \frac{1}{4\pi n_d m_d} B_z \frac{\partial B_z}{\partial x}.
    \label{eq:2}
\end{equation}
Here, $n_d$, $m_d$, $v_d$, and $Z_d$ represent the number density, mass, fluid velocity, and charge number of an individual dust grain, respectively. The elementary electronic charge is $e$, $\phi$ is the electrostatic potential, and $\psi$ represents the self-gravitational potential. Note that because of the large grain inertia ($m_d \gg m_i$) and low dust temperature ($T_d \ll T_i$), the dust thermal speed is negligible ($v_{thd} \ll v_{thi}$), allowing the dust thermal pressure gradient to be safely neglected in Eq.~\eqref{eq:2}.

To model the dynamic compression and stretching of the magnetic field lines perpendicular to the collapse direction, we adopt a single-fluid, ideal-MHD closure in which the magnetic field is frozen into the dust fluid. The induction equation then reads
\begin{equation}
    \frac{\partial B_z}{\partial t} + \frac{\partial}{\partial x}(v_d B_z) = 0,
    \label{eq:3}
\end{equation}
where $B_z$ is the dynamic magnetic field along the $z$-direction.

The electrostatic potential ($\phi$) and self-gravitational potential ($\psi$) are determined by the respective Poisson equations:
\begin{align}
    \frac{\partial^2 \phi}{\partial x^2} = 4\pi e (n_e - n_i + Z_d n_d),\label{eq:4}\\
    \frac{\partial^2 \psi}{\partial x^2} = 4\pi G m_d (n_d - n_{d0}).\label{eq:5}
\end{align}
Here, $n_e$ and $n_i$ denote the electron and ion number densities, $G$ is the universal gravitational constant, and $n_{d0}$ is the unperturbed equilibrium dust density. In Eq.~\eqref{eq:5}, the Jeans swindle is invoked by subtracting $n_{d0}$, ensuring a stable, infinite, and mathematically consistent homogeneous background at equilibrium (see Ref.~\cite{binneyGalacticDynamicsSecond2008}).

\subsection{Nonthermal $\kappa$ velocity distributions}
The lighter background species (electrons and ions) have thermal velocities far exceeding the characteristic dust flow velocity ($v_{the}, v_{thi} \gg v_d$). Due to ubiquitous acceleration processes in astrophysical plasma environments, these species are often characterized by high-energy power-law tails. We model their velocity profiles using the 1D isotropic $\kappa$-distribution (see Refs.~\cite{vasyliunasLowEnergyElectronsMagnetosphere1968,pierrardKappaDistributionsTheory2010,livadiotisKappaDistributionsTheory2017} for details), which has the form: 
\begin{align}
    f_{\kappa}(v_j) = &\frac{n_{j0}}{(\pi\kappa\theta^2_{j})^{\frac{1}{2}}}\frac{\Gamma(\kappa)}{\Gamma(\kappa-1/2)}\Bigg(1+\frac{v_j^2}{\kappa\theta^2_{j}} \Bigg)^{-\kappa}.\label{eq:6}
\end{align}
Here, $\theta_j = [(\kappa - 3/2)/\kappa]^{1/2} v_{\text{thj}}$ is the characteristic speed, where $v_{\text{thj}} = (k_B T_j / m_j)^{1/2}$ is the thermal velocity of species $j$ ($j = e, i$). The quantities $n_{j0}$,$v_j$, $m_j$, and $T_j$ represent the equilibrium number density, particle velocity, mass, and temperature of the $j$-th species, respectively. The standard gamma function is denoted by $\Gamma$ and $k_B$ is the Boltzmann constant. The spectral index $\kappa$ measures the degree of nonthermality and must exceed $3/2$ for $\theta_j$ to be finite. Smaller $\kappa\ (\to 3/2)$ values represent higher nonthermality, whereas for $\kappa\to\infty$, $f_\kappa$ reduces to the Maxwellian distribution at temperature $T_j$.

Integrating the distribution function in Eq.~\eqref{eq:6} over velocity space yields the self-consistent electron and ion densities \cite{pierrardKappaDistributionsTheory2010,livadiotisKappaDistributionsTheory2017}, respectively, as
\begin{align}
    n_e = n_{e0} \left[ 1 - \frac{e\phi}{k_B T_e (\kappa - 3/2)} \right]^{-(\kappa - 1/2)},\label{eq:7}\\
    n_i = n_{i0} \left[ 1 + \frac{e\phi}{k_B T_i (\kappa - 3/2)} \right]^{-(\kappa - 1/2)}.\label{eq:8}
\end{align}
Here, $n_{e0(i0)}$ and $T_{e(i)}$ are the equilibrium number density and temperature of electrons (ions). Again, as $\kappa \rightarrow \infty$, the densities revert to the Maxwell-Boltzmann thermal limits.

\subsection{Normalization scheme}
To eliminate raw physical constants, all the physical parameters are scaled using the unperturbed background ion parameters ($n_{i0}$, $T_i$, $m_i$) as the coordinate baseline. The resulting dimensionless variables are detailed in Table~\ref{tab:1}.
\begin{table}[htbp]
    \centering
    \caption{Normalization scheme used in the simulation.}
    \begin{ruledtabular}
        \begin{tabular}{ccc}
           Parameters & \hspace{-2em} Normalizing factors \hspace{1 em}& Normalized form \\ \hline \\
         Time  & $\omega_{pi}^{-1}$ & $\tilde{t} = \omega_{pi}t$ \\
         Space & $\lambda_{Di}$ & $\tilde{x} = x/\lambda_{Di}$ \\
         Velocity  & $v_{thi}$ & $\tilde{v}_d = v_d/v_{thi}$\\
         Density & $n_{i0}$ & $\tilde{n}_j = n_j/n_{i0}$\\
         Electrostatic potential & $k_B T_i/e$ & $\tilde{\phi} = e \phi/{k_B T_i}$\\
         Gravitational potential & $v_{thi}^2$ & $\tilde{\psi} = \psi/v_{thi}^2$\\
         Magnetic field & $B_\text{norm}$ & $\tilde{B}_z = B_z/B_\text{norm}$\\
        \end{tabular}
    \end{ruledtabular}
    \footnotesize{Here, $\omega_{pi} = (4\pi n_{i0}e^2/m_i)^{1/2}$, $\lambda_{Di} = (k_B T_i/ 4\pi n_{i0}e^2)^{1/2}$, $v_\text{thi} = (k_B T_i/m_i)^{1/2}$, and $B_\text{norm} = (n_{i0} k_B T_i)^{1/2}$.}
    \label{tab:1}
\end{table}

Applying this normalization scheme in Eqs.~\eqref{eq:1}--\eqref{eq:8} yields the following set of dimensionless equations:
\begin{gather}
\frac{\partial \tilde{n}_d}{\partial \tilde{t}}+\frac{\partial}{\partial \tilde{x}}(\tilde{n}_d\tilde{v}_d)=0, \label{eq:9}\\
\frac{\partial \tilde{v}_d}{\partial \tilde{t}}+\tilde{v}_d\frac{\partial \tilde{v}_d}{\partial \tilde{x}}= \frac{Z_d m_i}{m_d}\frac{\partial\tilde{\phi}}{\partial \tilde{x}}-\frac{\partial\tilde{\psi}}{\partial \tilde{x}} - \frac{m_i}{4 \pi m_d} \left(\frac{\tilde{B}_z}{\tilde{n}_d}\right) \frac{\partial \tilde{B}_z}{\partial \tilde{x}},\label{eq:10}\\
\frac{\partial \tilde{B}_z}{\partial \tilde{t}} + \frac{\partial}{\partial \tilde{x}}(\tilde{v}_d \tilde{B}_z) = 0,\label{eq:11}\\
\frac{\partial^{2}\tilde{\phi}}{\partial \tilde{x}^{2}}=\tilde{n}_e-\tilde{n}_i+Z_{d}\tilde{n}_d,\label{eq:12}\\
\frac{\partial^{2}\tilde{\psi}}{\partial \tilde{x}^{2}}=\alpha_G \left(\frac{\omega_{pd}}{\omega_{pi}}\right)^2\left(\frac{n_{i0}}{n_{d0}}\tilde{n}_d-1\right),\label{eq:13}\\
\tilde{n}_e=\left(\frac{n_{e0}}{n_{i0}}\right)\left[1-\frac{\tilde{\phi}T_i}{T_e(\kappa-3/2)}\right]^{-(\kappa-1/2)} \label{eq:14},\\
\tilde{n}_i=\left[1+\frac{\tilde{\phi}}{\kappa-3/2}\right]^{-(\kappa-1/2)}.\label{eq:15}
\end{gather}
Here, $\alpha_G = (\omega_{Jd}/\omega_{pd})^{2}$ is the relative self-gravity parameter, $\omega_{Jd} = (4 \pi G m_d n_{d0})^{1/2}$ is the dust Jeans frequency, and $\omega_{pd} = (4 \pi n_{d0} q_d^2 /m_d)^{1/2}$ represents the dust plasma frequency.

\subsection{Linear dispersion analysis}
Employing a plane-wave perturbation technique, with all perturbed
quantities varying as $\exp(i\tilde{k}\tilde{x}-i\tilde{\omega}\tilde{t})$, where $\tilde{k} = k\lambda_{Di}$ and $\tilde{\omega} = \omega/\omega_{pi}$, about the static ($\tilde{v}_{d0}=0$), homogeneous equilibrium, we linearize and Fourier transform Eqs.~\eqref{eq:9}--\eqref{eq:15} to obtain the following linear dispersion relation for the magnetized Jeans instability:
\begin{equation}
\tilde\omega^2 = \left(\frac{\omega_{pd}}{\omega_{pi}}\right)^2
\left[\frac{\tilde{k}^2}{\tilde{k}^2+\tilde\lambda_{eff}^{-2}}-\alpha_G\right]+ \left(\frac{v_A}{v_{\rm thi}}\right)^2 \tilde{k}^2.
\label{eq:16}
\end{equation}
Here, $\tilde\lambda_{eff}$ denotes the normalized effective nonthermal Debye length, defined as $\tilde\lambda_{\rm eff}=\lambda_{\rm eff}/\lambda_{Di}$, where $\lambda_{\text{eff}}=\left[k_B T_e T_i/4\pi e^2 \sigma_\kappa \left(n_{e0}T_i+n_{i0}T_e\right)\right]^{1/2}$ with $\sigma_\kappa = (\kappa -1/2)/(\kappa -3/2)$. The quantity $v_A$ corresponds to the Alfv\'en speed, given by $v_A=B_{z0}/\left(4\pi n_{d0} m_d\right)^{1/2}$.
In the unmagnetized limit ($B_{z0}\to0$), Eq.~\eqref{eq:16} reduces identically to the dispersion relation obtained in our previous works~\cite{ray_JSPC_2026,ray_PRE_2026}.

Equation~\eqref{eq:16} shows that the compressional magnetic term
$(v_A/v_{\rm thi})^2k^2$ enters additively and is independent of the nonthermal
spectral index $\kappa$, in contrast to the electrostatic screening
term, which is explicitly $\kappa$-dependent through $\tilde \lambda_{eff}$ (via $\sigma_\kappa$).
For instability ($\tilde\omega^2<0$), Eq.~\eqref{eq:16} requires $\alpha_G$
to exceed the magnetized critical threshold,
\begin{equation}
\alpha_{Gc}(\tilde{k}) = \frac{\tilde{k}^2}{\tilde{k}^2 + \tilde\lambda_{eff}^{-2}} + \left(\frac{\omega_{pi}}{\omega_{pd}}\right)^2 \left(\frac{v_A}{v_{\rm thi}}\right)^2\tilde{k}^2,
\label{eq:17}
\end{equation}
which increases monotonically with $v_A$ and hence $B_{z}$. This suggests that the ambient magnetic field raises the threshold for gravitational collapse through its added compressional pressure support.

\section{\label{sec:simulation setup} Simulation setup}
To simulate the nonlinear spatiotemporal evolution of the self-gravitating, magnetized dusty plasma, we employ a high-order, self-consistent 1D fluid simulation framework previously used in Ref.~\cite{ray_PRE_2026}. The normalized fluid and field equations are solved on a discretized spatial and temporal grid. Periodic boundary conditions are imposed across the simulation domain ($x \in [0, L_x]$) for all evolved fluid quantities ($n_d$, $v_d$, $B_z$) as well as the electrostatic and gravitational potentials ($\phi$ and $\psi$). Note that hereafter we drop the tilde symbol denoting normalized states as all the codes used are in normalized form. The first-order spatial derivatives in the dust continuity, momentum, and induction equations are approximated using a fourth-order accurate ($\sim\Delta x^4$, $\Delta x$ is the grid step) central finite-difference method \cite{omuraPlasmaWaveSignatures1993,kakadNonlinearEvolutionIon2014,singhFluidSimulationDustacoustic2022a} as follows:
\begin{align}
    \frac{\partial F_j^r}{\partial x} =& \frac{8 (F_{j+1}^r - F_{j-1}^r) - F_{j+2}^r + F_{j-2}^r}{12 \Delta x}, \label{eq:18}
\end{align}
where $F_j^r$ denotes any evolved field at grid index $j$ and temporal index $r$.

For second-order spatial derivatives in the electrostatic and gravitational Poisson equations, we use a fourth-order central finite-difference stencil \cite{omuraPlasmaWaveSignatures1993} as
\begin{align}
    \frac{\partial^2 F_j^r}{\partial x^2} =& \frac{16 (F_{j+1}^r + F_{j-1}^r) - 30 F_j^r - F_{j+2}^r - F_{j-2}^r }{12 (\Delta x)^2}.\label{eq:19}
\end{align}

Temporal advancement is performed with the explicit leap-frog scheme, which is second-order accurate in time ($\sim \Delta t^2$, $\Delta t$ being the time step) \cite{kakadFirsteverModelSimulation2016, lotekarGenerationIonAcoustic2017,singhFluidSimulationDustacoustic2022a}:
\begin{align}
    \frac{\partial F_j^r}{\partial t} = \frac{F_j^{r+1} - F_j^{r-1} }{2\Delta t}.\label{eq:20}
\end{align}

The lighter species (electrons and ions) respond to the potential on timescales much shorter than those of the dust dynamics ($v_{the},v_{thi}\gg v_d$). They therefore have no evolution equations of their own, and their densities adapt instantaneously to $\phi$ through Eqs.~\eqref{eq:14} and \eqref{eq:15}. Because $n_e$ and $n_i$ depend nonlinearly on $\phi$, the electrostatic Poisson equation must be solved iteratively at every time step, and the gravitational Poisson equation is solved in the same way. We use successive over-relaxation (SOR) with the five-point stencil of Eq.~\eqref{eq:19}, following Refs.~\cite{ray_PRE_2026,lotekarEffectiveApproachImplement2019}. In contrast to our previous study~\cite{ray_PRE_2026}, which used the three-point second-order stencil, this keeps the spatial order of the Poisson solves equal to that of the fourth-order derivatives in Eq.~\eqref{eq:18}. Denoting the iteration index within a time step by $s$, the discrete updates for Eqs.~\eqref{eq:12} and \eqref{eq:13} are
\begin{align}
    \phi_j^{s+1} =& (1-\zeta)\phi_j^{s} + \frac{\zeta}{30} \bigg[16\left(\phi_{j-1}^{s+1} + \phi_{j+1}^{s}\right) - \phi_{j+2}^{s} \nonumber \\& - \phi_{j-2}^{s+1} - 12\Delta x^2 (n_e - n_i + Z_d n_d)\bigg],\label{eq:21}\\
    \psi_j^{s+1} =& (1-\zeta)\psi_j^{s} + \frac{\zeta}{30} \bigg[16\left(\psi_{j-1}^{s+1} + \psi_{j+1}^{s}\right) - \psi_{j+2}^{s} \nonumber \\& - \psi_{j-2}^{s+1} - 12\Delta x^2 \alpha_{G} \left(\frac{\omega_{pd}}{\omega_{pi}}\right)^2 \left(\frac{n_{i0}}{n_{d0}}n_d - 1\right)\bigg].\label{eq:22}
\end{align}
Here, $\zeta$ is the relaxation parameter. In Eq.~\eqref{eq:21}, $n_e$ and $n_i$ are re-evaluated from Eqs.~\eqref{eq:14} and \eqref{eq:15} with the current iterate of $\phi$ at every sweep, whereas $n_d$ is held at its value at the current time step. The nodes $j-1$ and $j-2$ use the newly computed values $s+1$, and the nodes $j+1$ and $j+2$ use the values from iteration $s$. The potentials from the preceding time step serve as the initial guess, and the iterations are terminated when $|\phi_j^{s+1}-\phi_j^{s}|<\eta$ and $|\psi_j^{s+1}-\psi_j^{s}|<\eta$ ($\eta$ being the convergence tolerance) for the electrostatic and gravitational Poisson equations, respectively.

Leap-frog schemes often exhibit a grid-separation instability that generates high-frequency grid-level noise. To suppress these numerical short-wavelength discretization errors, we pass the updated fluid and field variables through a fourth-order compensated low-pass filter at the end of every time step \cite{kakadNonlinearEvolutionIon2014,lotekarFluidSimulationDispersive2016,lotekarEffectiveApproachImplement2019}:
\begin{align}
    F_j^* =& \frac{5}{8} F_j + \frac{1}{4}(F_{j-1} + F_{j+1}) - \frac{1}{16}(F_{j-2}+F_{j+2}).
    \label{eq:23}
\end{align}
The filter operates as a localized smoother in Fourier space, dampening unphysical grid-level modes while leaving the physically meaningful, long-wavelength self-gravitational Jeans waves completely unattenuated.

The discretization parameters $\Delta x$ and $\Delta t$ are chosen to satisfy the Courant--Friedrichs--Lewy (CFL) stability constraint \cite{birdsallPlasmaPhysicsComputer2004}, $v_{\text{max}}\left(\Delta t / \Delta x\right) \le 1$, where $v_{\text{max}}$ is the fastest characteristic propagation velocity in the system.

To trigger the instability and observe structural growth, a localized Gaussian density perturbation of the following form is introduced at the center of the domain \cite{lotekarGenerationIonAcoustic2017,singhFluidSimulationDustacoustic2022a}:
\begin{align}
        \delta n = \Delta n \exp \left[-\left(\frac{x-x_c}{l_0}\right)^2\right],\label{eq:24}
\end{align}
where $\Delta n$ is the perturbation amplitude, $l_0$ is the spatial perturbation width, and $x_c$ represents the center of the domain.

For all simulation runs, we select a fixed grid spacing of $\Delta x = 0.03$ and a time step of $\Delta t = 1.2 \times 10^{-4}$ on a grid of $N_x = 8192$ points, resulting in a box length of $L_x = N_x \Delta x = 245.76\,\lambda_{Di}$. The Poisson solver convergence tolerance is set to $\eta = 10^{-5}$ with a relaxation parameter of $\zeta = 1.2$. The initial relative density perturbation amplitude is $\Delta n = 0.02 \times 10^{-3}$ with a spatial width of $l_0 = 100$ grid points. Based on the ion-baseline normalization ($n_{i0} = 1$), the unperturbed dust number density is set to $n_{d0} = 10^{-3}n_{i0}$ with a dust charge number of $Z_d = 100$, which yields an unperturbed electron number density of $n_{e0} = n_{i0} - Z_d n_{d0} = 0.9n_{i0}$ via the quasi-neutrality condition. The other input parameters are $m_d = 1000m_i$, $m_e = (1/1836)m_i$, $T_d = 0.1T_i$, and $T_e = 10T_i$. Using these input parameters, we obtain $\omega_{pd} = 0.1\omega_{pi}$, $\omega_{pe} = 40.65\omega_{pi}$, $\lambda_{Dd} = 0.1\lambda_{Di}$, and $\lambda_{De} = 3.33\lambda_{Di}$. The dust and electron thermal speeds are $v_{thd} = 0.01v_{thi}$ and $v_{the} = 135.49v_{thi}$, respectively. Furthermore, the $B_{z}$ value is varied in the range $B_{z} = 0 - 2 \ B_{\rm norm}$, while $\kappa$ is varied as $\kappa = 2,5, 20, \text{ and }100$. The relative self-gravity parameter $(\alpha_G)$ is treated as a free control parameter. For the fundamental simulation mode ($k=2\pi/L_x=0.0256$) and $\kappa=5$, Eq.~\eqref{eq:17} yields $\alpha_{Gc}\approx4.7\times10^{-4}$ at $B_z=0$, increasing to $\alpha_{Gc}\approx2.1\times10^{-2}$ at the largest field considered ($B_z=2$). Thus, a fixed value of $\alpha_G=0.1$ is adopted in all simulation runs, which remains well above threshold for the longest-wavelength modes across the entire range of $B_z$ explored here, ensuring the system remains unstable to gravitational collapse. A similar margin holds for the other values of $\kappa$ examined in this study. The parameter sets used in each run are summarized in Table~\ref{tab:2}.
\begin{table}[htbp]
    \centering
    \caption{Input parameters for different simulation runs.}
    \begin{ruledtabular}
        \begin{tabular}{lccc}
            Run & Sub-run &$\kappa$ & $B_{z}[B_{\rm norm}]$\\
            \colrule
            \multirow{5}{*}{R1} & i & \multirow{5}{*}{2} & 0 \\
            & ii &  & 0.4\\
            & iii &  & 0.8\\
            & iv &  & 1.2\\
            & v &  & 2.0\\
            \colrule
            \multirow{5}{*}{R2} & i & \multirow{5}{*}{5} & 0 \\
            & ii &  & 0.4\\
            & iii &  & 0.8\\
            & iv &  & 1.2\\
            & v &  & 2.0\\
            \colrule
            \multirow{5}{*}{R3} & i & \multirow{5}{*}{20} & 0 \\
            & ii &  & 0.4\\
            & iii &  & 0.8\\
            & iv &  & 1.2\\
            & v &  & 2.0\\
            \colrule
            \multirow{5}{*}{R4} & i & \multirow{5}{*}{100} & 0 \\
            & ii &  & 0.4\\
            & iii &  & 0.8\\
            & iv &  & 1.2\\
            & v &  & 2.0\\
        \end{tabular}
    \end{ruledtabular}
    \label{tab:2}
    \footnotesize{\textbf{Note.} The fixed simulation parameters across all the runs are: $\Delta x = 0.03$, $\Delta t = 1.2 \times 10^{-4}$, $\Delta n = 0.02 \times 10^{-3}$, $l_0 = 100$, $n_{d0} = 10^{-3}n_{i0}$, $n_{e0} = 0.9n_{i0}$, $Z_d = 100$, $m_d = 1000m_i$, $m_e = (1/1836)m_i$, $T_d = 0.1T_i$, and $T_e = 10T_i$}.
\end{table}

The simulations are dimensionless, so a given run applies to any plasma with the same dimensionless parameters. The physical units follow as the ion-baseline is chosen. We take protons as the ion species with $n_{i0}=1~{\rm cm^{-3}}$, $T_i=1$~eV, $T_e=10$~eV, $T_d=0.1$~eV, $n_{d0}=10^{-3}~{\rm cm^{-3}}$ and $Z_d=100$, which gives $\omega_{pi}=1.32\times10^{3}~{\rm rad\,s^{-1}}$, $\lambda_{Di}=7.43\times10^{2}$~cm, $v_{thi}=9.8\times10^{5}~{\rm cm\,s^{-1}}$ and $B_{\rm norm}=(n_{i0}k_BT_i)^{1/2}=1.27~\mu$G. Thus, the scanned fields ($B_z=0,\,0.4,\,0.8,\,1.2,\,2.0\,B_{\rm norm}$) correspond to $0,\,0.51,\,1.02,\,1.52,\,2.54~\mu$G. Zeeman observations of the diffuse interstellar medium give field strengths of a few to $\sim10~\mu$G, with a median of $\approx6~\mu$G in the cold neutral medium~\cite{crutcherMAGNETICFIELDSINTERSTELLAR2010,heilesMillenniumArecibo212005}. The scan therefore covers the weak to moderate part of the observed range.

\section{\label{sec:results and discussion}Results and discussion}
We have used a total of 20 simulation runs for different combinations
of the magnetic field strength $(B_z)$ and the nonthermal spectral
index $(\kappa)$ at a fixed value of the relative self-gravity
parameter $(\alpha_G)$, as described in Sec.~\ref{sec:simulation setup}. 

\subsection{Spatiotemporal evolution}
\begin{figure*}
    \centering
    \includegraphics[width=\linewidth]{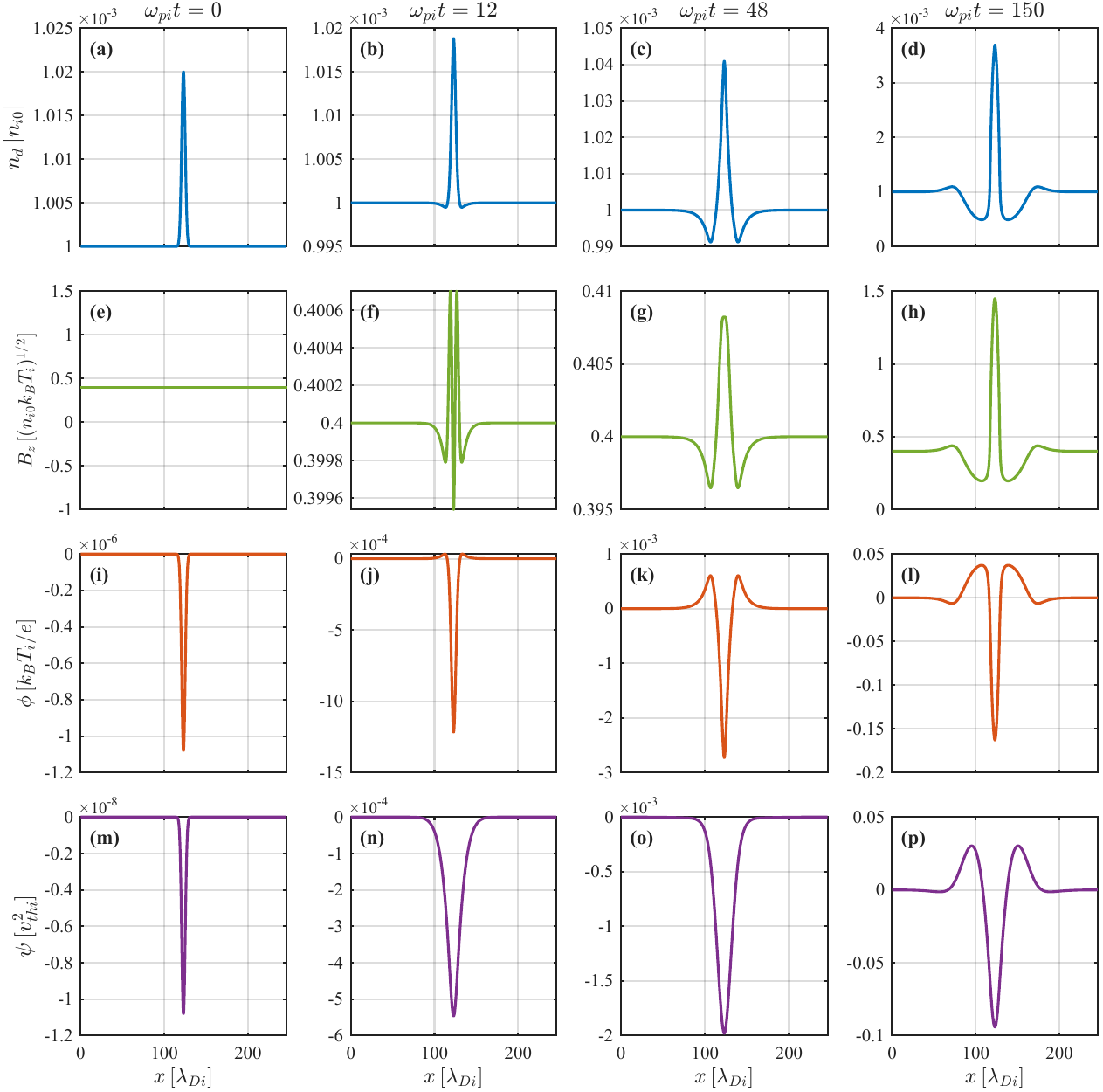}
    \caption{Spatiotemporal evolution of normalized dust density $n_d[n_{i0}]$, magnetic field $B_z [(n_{i0}k_BT_i)^{1/2}]$, electrostatic potential $\phi [k_BT_i/e]$, and gravitational potential $\psi[v_{thi}^2]$ for $\kappa = 5$ and $B_z = 0.4\ B_{\rm norm}$ at $\alpha_{G} = 0.1$.}
    \label{fig:1}
\end{figure*}

Figure~\ref{fig:1} shows the nonlinear spatiotemporal evolution of the normalized dust density $n_d [n_{i0}]$, perpendicular magnetic field $B_z [(n_{i0}k_BT_i)^{1/2}]$, electrostatic potential $\phi [k_BT_i/e]$, and gravitational potential $\psi [v_{thi}^2]$ for a representative magnetized run ($\kappa=5$, $B_{z}=0.4B_{\rm norm}$, $\alpha_G=0.1$) at four different time steps. At $\omega_{pi}t = 0$, the system is initialized with a localized, small-amplitude ($\Delta n = 0.02 \times 10^{-3}$, $2\%$ of $n_{d0}$) Gaussian density perturbation centered in the simulation domain (Fig.~\ref{fig:1}a). This initial concentration of mass and negative charge immediately establishes sharp, highly localized electrostatic and gravitational potential wells (Fig.~\ref{fig:1}i and \ref{fig:1}m), while the ambient magnetic field remains uniform at an initial strength of $B_{z} = 0.4 B_{\rm norm}$ (Fig.~\ref{fig:1}e). During the early transient phase ($\omega_{pi}t = 0 - 12$), the system undergoes an electrostatic decompression. The dense core of negatively charged dust generates an outward electrostatic pressure that initially overpowers the inward gravitational pull. This outward fluid motion drives mass away from the center, resulting in a decrease of the central density peak and the formation of distinct density rarefaction troughs symmetrically adjacent to the central peak. Consequently, $\phi$ broadens, developing positive potential shoulders that flank the central negative well. Meanwhile, the gravitational potential $\psi$ deepens and broadens (Fig.~\ref{fig:1}n) as the mass perturbation spreads and redistributes across the central region. As the magnetic field lines are frozen into the dust fluid, this early outward expansion displaces the magnetic field, generating fine, oscillatory perturbations in $B_z$ (Fig.~\ref{fig:1}f). Physically, this transient broadening reflects the dispersive propagation of the stable, short-wavelength Fourier components of the initial Gaussian perturbation, which radiate outward before the unstable long-wavelength modes take over.

As the system continues to evolve ($\omega_{pi}t = 12 - 48$), the self-gravitational pull halts the outward electrostatic expansion and reverses the fluid flow. The central density begins to re-accumulate mass, deepening the adjacent rarefaction troughs (Fig.~\ref{fig:1}c). Alongside this inward fluid reversal, the magnetic field perturbation consolidates into a distinct central compression peak perfectly aligned with the density core, flanked by deepening magnetic depletion regions (Fig.~\ref{fig:1}g). Furthermore, the positive shoulders of the electrostatic potential become highly pronounced as the charge separation intensifies (Fig.~\ref{fig:1}k), while the gravitational potential well continues to intensify (Fig.~\ref{fig:1}o), reinforcing the inward accretion as central mass concentration grows.

By $\omega_{pi}t = 150$, the system has entered the nonlinear runaway regime. Rather than a single localized bump, all four fields now display a broad, strongly modulated region flanking a sharp central spike. The dust density peak has grown to $\sim 3.6\,n_{d0}$, while the surrounding rarefaction flanks have dropped to less than half of $n_{d0}$ (Fig.~\ref{fig:1}d). Correspondingly, the electrostatic potential well and its flanking positive shoulders amplify markedly (Fig.~\ref{fig:1}l), while the gravitational potential $\psi$ deepens substantially and develops positive shoulders driven by the local mass deficit ($n_d < n_{d0}$) in the depleted density flanks (Fig.~\ref{fig:1}p). Crucially, the ideal magnetohydrodynamic flux-freezing condition forces the transverse magnetic field lines to be dragged inward and compressed alongside the collapsing dust fluid. This coupling produces a pronounced localized amplification of the magnetic field, which peaks at approximately $1.5\,B_{\rm norm}$, nearly four times its initial strength (Fig.~\ref{fig:1}h). This synchronized spatiotemporal compression illustrates the buildup of the magnetic-pressure response; as self-gravity contracts the dust fluid transverse to the field lines, $B_z$ is locally amplified and establishes an outward magnetic-pressure gradient. Together, these snapshots trace the progression from early transient wave dispersion and linear amplification to pronounced nonlinear structure formation.

\subsection{Effects of magnetic field}
\begin{figure}
    \centering
    \includegraphics[width=\linewidth]{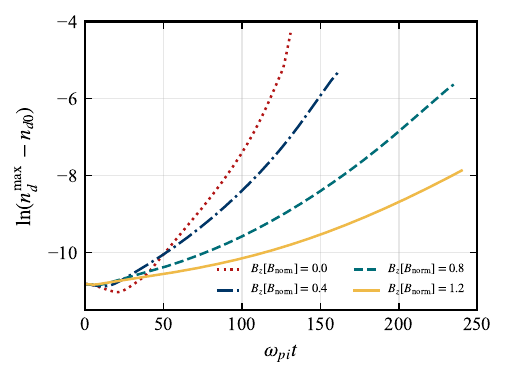}
    \caption{Temporal evolution of the natural logarithm of the maximum dust density perturbation $\ln(n_{d}^{\max} - n_{d0})$ for different $B_z$ at $\kappa=5$.}
    \label{fig:2}
\end{figure}

To quantify the influence of the magnetic field on nonlinear growth stage of the Jeans instability, Fig.~\ref{fig:2} illustrates the time evolution of the natural logarithm of the maximum dust density perturbation $\ln(n_{d}^{\max} - n_{d0})$ for different magnetic field strengths ($B_{z} [B_{\rm norm}] = 0.0, 0.4, 0.8, 1.2$) at $\kappa = 5$ and $\alpha_G = 0.1$.

At early times ($\omega_{pi}t \sim 20$), all cases display a transient reduction in the central dust density, corresponding to the electrostatic decompression phase seen in Fig.~\ref{fig:1}. However, introducing an ambient magnetic field significantly flattens this density dip. Because the field is frozen into the dust fluid, any transverse displacement carries the field lines with it, generating an opposing magnetic-pressure response. Consequently, magnetic pressure resists the initial outward electrostatic expansion, limiting the amplitude of the early density dip.

Following this brief electrostatic decompression, self-gravity overcomes electrostatic repulsion and reverses the flow, leading to sustained mass accumulation in the core. In this subsequent stage, the profiles exhibit an approximately linear increase in logarithmic space, marking the onset of exponential growth. The simulated nonlinear growth rate $\gamma_{\rm sim}$ is extracted directly from the slopes of these curves by applying a linear regression to $\ln(n_d^{\max} - n_{d0})$ during the exponential phase, yielding coefficients of determination $R^2 \ge 0.997$ across all 20 simulation runs. The temporal fitting windows are selected to isolate these linear segments while avoiding both the early transient bounce and late-stage nonlinear saturation. The resulting values of $\gamma_{\rm sim}$ for all $\kappa$ and $B_z$ are summarized in Table~\ref{tab:3}. A systematic attenuation of the exponential growth with increasing $B_z$ is clearly evident. The unmagnetized case ($B_z = 0$) exhibits the steepest growth, whereas increasing $B_z$ up to $1.2\,B_{\rm norm}$ progressively flattens the slope. Consistent with the flux-frozen core compression observed in Fig.~\ref{fig:1}, stronger ambient fields build up a larger outward magnetic-pressure barrier that opposes inward transverse contraction in the same way it resisted the initial outward expansion, directly retarding gravitational collapse and prolonging the lifetime of the localized structure prior to runaway.

\begin{table}[htbp]
    \centering
    \begin{ruledtabular}
        \begin{tabular}{ccccc}
        \textrm{Sl. No.}&
        $\kappa$& $B_z [B_{\rm norm}]$&
        $\gamma_\text{sim} [\omega_{pi}]$& Fitting window\\
        \colrule
         1 & \multirow{5}{*}{2}& 0.0 & 0.0542 & 35 to 85\\
         2 & & 0.4 & 0.0273 & 20 to 70 \\
         3 & & 0.8 & 0.0129 & 20 to 70\\
         4 & & 1.2 & 0.0067 & 20 to 70\\
         5 & & 2.0 & 0.0024 & 20 to 70\\
         \colrule
         6 & \multirow{5}{*}{5}& 0.0 & 0.0498 & 40 to 90\\
         7 & & 0.4 & 0.0265 & 20 to 70\\
         8 & & 0.8 & 0.0123 & 20 to 70\\
         9 & & 1.2 & 0.0065 & 20 to 70\\
         10 & & 2.0 & 0.0024 & 20 to 70\\
         \colrule
         11 & \multirow{5}{*}{20}& 0.0 & 0.0466 & 40 to 90\\
         12 & & 0.4 & 0.0246 & 20 to 70\\
         13 & & 0.8 & 0.0121 & 20 to 70\\
         14 & & 1.2 & 0.0064 & 20 to 70\\
         15 & & 2.0 & 0.0024 & 20 to 70\\
         \colrule
         16 & \multirow{5}{*}{100}& 0.0 & 0.0460 & 45 to 90\\
         17 & & 0.4 & 0.0244 & 20 to 70\\
         18 & & 0.8 & 0.0121 & 20 to 70\\
         19 & & 1.2 & 0.0064 & 20 to 70\\
         20 & & 2.0 & 0.0024 & 20 to 70\\
        \end{tabular}
    \end{ruledtabular}
    \caption{Simulation growth rate $\gamma_{\text{sim}}$ for different values of $\kappa$ and $B_z$ at $\alpha_G = 0.1$. All linear regressions yield $R^2 \ge 0.997$.}
    \label{tab:3}
\end{table}

\begin{figure}
    \centering
    \includegraphics[width=\linewidth]{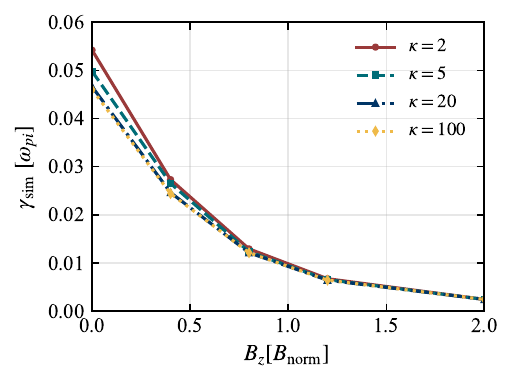}
    \caption{Variation of the simulated growth rate $\gamma_\text{sim} [\omega_{pi}]$ as a function of $B_z$ for different $\kappa$. The red solid line, cyan dashed line, blue dashed-dotted line, and yellow dotted line correspond to $\kappa = 2, 5, 20, \text{ and } 100$, respectively.}
    \label{fig:3}
\end{figure}

Figure~\ref{fig:3} depicts $\gamma_{\rm sim}[\omega_{pi}]$ as a function of $B_{z}[B_{\rm norm}]$ for different $\kappa$ at $\alpha_G=0.1$. For all $\kappa$, $\gamma_{\rm sim}$ decreases monotonically with increasing $B_z$, dropping from $\gamma_{\rm sim}\approx 0.046$--$0.054\,\omega_{pi}$ at $B_z=0$ to $\gamma_{\rm sim}\approx 0.0024\,\omega_{pi}$ at $B_z=2.0$, a reduction exceeding one order of magnitude. In addition, the $\gamma_{\rm sim}$ curves systematically converge as $B_z$ increases, indicating a progressive reduction of the $\kappa$-dependent variation in the growth rate. The largest separation occurs in the unmagnetized and weakly magnetized regimes, where the most nonthermal case ($\kappa=2$) yields the highest growth rate; this separation diminishes with increasing $B_z$ as magnetic compression provides the dominant stabilizing contribution. At $B_z\simeq 2$, the growth rates for all $\kappa$ become nearly identical, consistent with magnetic restoring forces overwhelming the distribution-dependent electrostatic/kinetic effects. This simulated macroscopic behavior follows directly from the analytical structure of the dispersion relation, Eq.~\eqref{eq:16}. The spectral index $\kappa$ influences the growth rate through the effective shielding length $\lambda_{\rm eff}$ (via $\sigma_\kappa$), modifying the electrostatic term $k^2/(k^2+\lambda_{\rm eff}^{-2})$. However, the compressional magnetic term $(v_A/v_{\rm thi})^2k^2$ is entirely independent of $\kappa$. At weak fields, the $\kappa$-dependent screening term dominates the restoring force, and the sensitivity to nonthermality is visible in $\gamma_{\rm sim}$. As $B_z$ increases, the $\kappa$-independent magnetic term grows in relative magnitude until it overwhelmingly dominates the total restoring pressure. Once this threshold is crossed, the growth rate, both analytically in the linear regime and empirically in the fully nonlinear simulations, becomes effectively insensitive to the velocity distribution of the lighter plasma species. Physically, this demonstrates that in sufficiently magnetized environments, the thermodynamic state of the background plasma is entirely subdominant to the ambient magnetic field in regulating the onset of structural collapse. Finally, the near overlap of the $\kappa=20$ and $\kappa=100$ curves across all $B_z$ is consistent with the approach to the Maxwellian limit at large values of $\kappa$.

\section{\label{sec:conclusions}Conclusions}
In this study, we have investigated how magnetic fields and nonthermal particle statistics jointly regulate nonlinear gravitational collapse in a magnetized, tricomponent complex plasma. The lighter nonthermal electrons and ions follow a $\kappa$-distribution, while the inertial dust is coupled to a uniform ambient magnetic field, frozen into the dust and perpendicular to the direction of variation. We derived the linear dispersion relation of the magnetized system followed by an extensive suite of nonlinear fluid simulations. The spatiotemporal evolution of different plasma parameters is thoroughly examined to quantify the effects of a transverse magnetic field on a nonthermal self-gravitating system and its subsequent collapse. The main findings are summarized as follows:
\begin{enumerate}
    \item The magnetic field slows gravitational collapse. In the linear theory, the field adds a compressional term to the dispersion relation that raises the critical self-gravity threshold and narrows the band of unstable wavelengths. In the simulations, the collapsing core compresses the frozen-in field, and the resulting magnetic pressure gradient opposes gravity, consistently with the linear picture. The growth rate decreases monotonically and strongly with increasing strength of the magnetic field for every $\kappa$ considered.
    \item The influence of nonthermality over the instability decreases as the strength of the magnetic field increases. In unmagnetized and weakly magnetized environments, enhanced nonthermality (smaller $\kappa$) gives a higher growth rate, because the $\kappa$-dependent electrostatic screening controls the response. However, as the magnetic term is independent of $\kappa$, the growth rates of thermal and nonthermal plasmas progressively converge as the field strength increases, and in sufficiently magnetized environments, the growth rates become largely insensitive to the degree of nonthermality. Thus, in such a strong-field regime, magnetic pressure, rather than the velocity distribution of the lighter species, governs the collapse rate.
\end{enumerate}
These results generalize our earlier unmagnetized study \cite{ray_PRE_2026} and demonstrate that even a moderate transverse magnetic field can provide substantial stabilizing support, thereby delaying or suppressing Jeans-instability-driven collapse in nonthermal dusty plasmas. This carries direct implications for structure formation in magnetized, nonthermal environments such as diffuse interstellar clouds, where both effects are simultaneously present. However, it should be noted that the model considered here is one-dimensional with a uniform magnetic field perpendicular to the direction of collapse. Consequently, the stabilization applies to contraction across the field, while contraction along the field lines is unaffected. This transverse geometry isolates compressional magnetosonic support but does not capture the more general case of oblique field orientations expected in star-forming regions. Furthermore, the ideal frozen-in approximation neglects non-ideal dissipative processes, such as ambipolar diffusion, Ohmic resistivity, and dust-neutral collisional drag. Finally, the dust component is treated as a cold, fixed-charge fluid, omitting dust thermal pressure, grain-size distributions, and self-consistent dynamical charging. Extending the present framework to oblique and multi-dimensional configurations with non-ideal effects  and dust charging constitutes a natural direction for future investigations.

\section*{Acknowledgments}
We express our sincere gratitude to Tezpur University for the invaluable support and resources. A special appreciation is extended to all the members of the Astrophysical Plasma and Nonlinear Dynamics Research Laboratory (APNDRL) for the insightful discussions. D.R. acknowledges the partial financial support received from Tezpur University, Assam, India, vide sanction order no. F. 11-105 VI(E)/3056 (dated 11.11.2026). A.K. and B.K. acknowledge support from DST/IIG/MIPEARL/2024-2027 project. P.K.K. is thankful for the academic associateship by Inter-University Centre for Astronomy and Astrophysics (IUCAA), Pune, India.

\section*{Data Availability}
The data generated and used in this investigation are available upon reasonable request from the corresponding author. 

\bibliography{apssamp}

\end{document}